\documentclass[12pt,letterpaper]{amsart}
\usepackage{amssymb}
\usepackage[dvipdfmx]{graphicx}
\title[Chiral algebras' anomaly embedded to smooth surfaces]{Mixed anomalies of chiral algebras compactified to smooth quasi-projective surfaces}

\author{Makoto Sakurai}
\date{February 25, 2015}


\begin{document}

\begin{abstract}

Some time ago, the chiral algebra theory of Beilinson-Drinfeld\cite{BD} was expected to play a central role in the convergence of divergence in mathematical physics of superstring theory for quantization of gauge theory and gravity.
Naively, this algebra plays an important role in a holomorphic conformal field theory with a non-negative integer graded conformal dimension, whose target space does not necessarily have the vanishing first Chern class.
This algebra has two definitions until now: one is that by Malikov-Schechtman-Vaintrob by gluing affine patches, and the other is that of Kapranov-Vasserot by gluing the formal loop spaces.
I will use the new definition of Nekrasov by simplifying Malikov-Schechtman-Vaintrob in order to compute the obstruction classes of gerbes of chiral differential operators.

In this paper, I will examine the two independent Ans\"{a}tze (or working hypotheses) of Witten's $\mathcal{N}=(0,2)$ heterotic strings and Nekrasov's generalized complex geometry, after Hitchin and\\
Gualtieri, are consistent in the case of $\mathbb{CP}^2$, which has $3$ affine patches and is expected to have the ``first Pontryagin anomaly".

I also scrutinized the physical meanings of $2$ dimensional toric Fano manifolds, or rather toric del Pezzo surfaces, obtained by blowing up the non-colinear $1, 2, 3$ points of $\mathbb{CP}^2$.
The obstruction classes of gerbes of them coincide with the second Chern characters obtained by the Riemann-Roch theorem and in particular vanishes for $1$ point blowup, which means that one of the gravitational anomalies vanishes for a non-Calabi-Yau manifold compactification.

The future direction towards the geometric Langlands program is also discussed in the last section.
\end{abstract}
\maketitle
\newpage
\textbf{AMS Mathematics Subject Classification Numbers:}\\
\textbf{14D21}Applications of vector bundles and moduli spaces in mathematical physics (twistor theory, instantons, quantum field theory),\\
\textbf{14F43}Other algebro-geometric (co)homologies (e.g., intersection, equivariant, Lawson, Deligne (co)homologies),\\
\textbf{14J26} Rational (or ruled) surfaces,\\
\textbf{55N05} Cech types,\\
\textbf{81T40} Two-dimensional field theories, conformal field theories, etc.
\tableofcontents
\newpage
\section{Introduction}
The chiral algebra of Beilinson and Drinfeld \cite{B}\cite{BF}\cite{Gaitsgory} \footnote{Some authors call it a ``pull-back functor".}from world-sheet viewpoints as well as chiral de Rham complex (factorization algebra) of Malikov, Schechtman, and Vaintrob \cite{MSV} from the target space viewpoints, are attempts to extend rational conformal field theory\cite{MS} (CFT of topological sigma model) to the case of target manifolds which do not necessarily obey the condition of the vanishing first Chern class of tangent bundle $c_1=0$.
We usually exclude such a situation from our consideration since a non-linear sigma model on such a target manifold contains a non-vanishing beta function.
Scale invariance will be lost in such a situation and one can not apply the method of CFT.

One rephrasing of the idea of Beilinson and Drinfeld is to use the patchwise construction by Malikov-Schectman-Vaintrob\cite{MSV} where one considers CFT defined on each coordinate patch and then considers the consistency of the theory under coordinate transformations among different patches. There is another \underline{coordinate-free definition} of chiral de Rham complex by Kapranov-Vasserot(see \cite{KV}\cite{KV4}) elaborating the motivic integration over formal loop / arc spaces for the family of maps from Riemann surfaces to a complex algebraic manifold, but I do not utilize this definition in this paper because its non-linear chiral algebra (factorization algebra) demands too much of ``abstract nonsense" category theory and it is not capable of comparing it to a geometric realization of its original definition due to Kapranov and Vasserot to the higher Chern character theory without assuming any unreliable groundless axioms of stochastics.

In the case when the manifold is covered by, say, $4$ patches $U_0, U_1, U_2$,\\
$U_3$, one first considers successive transformations $U_i\rightarrow U_{i+1}$.
In the end one finds that under the total coordinate change $U_0\rightarrow U_1 \rightarrow U_2\rightarrow U_3\rightarrow U_0$ the fields do not quite come back to their original values but there appears an additional term.
Namely, there exists an obstruction or anomaly for a consistent CFT in such a system described by a sheaf (or more abstractly by gerbe) cohomology\cite{KS}. It has been suggested, or even proved in cases of \underline{``differentiable varieties"} in the broad sense of K\"ahler manifolds (not assuming the existence of generalized complex structures) as the target spaces, that the obstruction is related to the first Pontryagin class (or rather, the second Chern character of non-negative grading of creation operators) of the dg-manifold by Fontaine-Kapranov\cite{FK}, Gorbounov-Malikov-Schechtman\cite{GMS},\\
Kapranov-Vasserot\cite{KV2}, Nekrasov\cite{N}, and Witten\cite{Witten}.

In this paper, I simplify the computational approach of Malikov-Schechtman-Vaintrob (see \cite{MSV}) and study the case of del Pezzo surfaces which are rational surfaces obtained from $\mathbb{CP}^2$ by blowing up a certain number of each points by co-dimension $1$ submanifold(s) (divisor(s)).
I study the cases of one-point, two-point, and three-point blowups of $\mathbb{CP}^2$ and show that the anomaly on these surfaces is in fact proportional to the first Pontryagin class (or rather; the second Chern character).
I consider this is a substantial simplification of the computation by Witten and Nekrasov by $\beta \gamma$ ghosts, in terms of cotangent bundles, equivalent to the topological half-twist of supersymmetry algebras.
In section $2$ of this paper I start from the general theory of $\beta, \gamma$
system (conformal dimensions of $\gamma, \beta$ are $0$ and $1$,
respectively) where $\gamma$ field is identified as the local coordinate of the manifold and $\beta$ field is identified as an $1$-form.
Following Malikov-Schechtman-Vaintrob(see \cite{MSV}) and Nekrasov\cite{N}, I discuss the transformation laws of $\gamma,\beta$ system under coordinate change so that their OPE (Operator-Product-Expansion) is preserved. I will introduce the anomaly (holomorphic closed) $2$-form of Nekrasov\cite{N} in order to simplify the computation of OPE.
Then in section $3$, I discuss the case of $\mathbb{CP}^2$ as the target manifold and reproduce the result of Witten.
In section $4$, I consider the case of del Pezzo surfaces (not including Hirzebruch surfaces or Enriques surfaces) from the $1, 2, 3$-point blowup of $\mathbb{CP}^2$ and no Pontryagin anomaly occurs in the $1$ point blow-up.
I computed the obstruction for these cases and find that they are proportional to the first Pontryagin class (or, the second Chern character) of the manifolds.
Computations become simplified after the topological halt-twist.
The non-trivial aspects of computation are changes of the convention for normal ordering when one goes to a different coordinate patch and one has to make a careful analysis.
In section $5$, I present some discussions and conclusions.

\subsection{Definition of del Pezzo surfaces and\\the second Chern character from Noether's formula}

\quad Del Pezzo surfaces are defined (in this paper) as the irreducible smooth quasi-projective proper, but not a priopi compact (in the mixed Hodge theory sense), complex algebraic surfaces $X$ whose anticanonical divisor is ample. In this paper, the anticanonical divisor of $X$ is
\begin{eqnarray}
-K_X & = & 3 H - E_1 - \cdots - E_n, \nonumber\\
H^2 & = & 1, \nonumber\\
E_i . E_j & = & - \delta_{ij}, \nonumber\\
H . E_i & = & 0, \nonumber
\end{eqnarray}
where $H$ is the hyperplane class and $E_i (i = 1, \cdots, n)$ is the exceptional divisor of $0 \le n \le 8$, but let us assume $n \leq 3$ for the toric diagrams, for the sake of generic point(s) blowups of $\mathbb{CP}^2$. As such, I do not deal with singular del Pezzo (and orbifolds) or simultaneous blowing up by $(-2)$-curves of divisor class $E_i$ like $E_i . E_i = -2$. Therefore one obtains $K_X^2 = 9 - n$. I will treat the generic $n$ $(n = 0, 1, 2, 3)$ point blowups of $\mathbb{CP}^2$ by attaching $\mathbb{CP}^1$ instead of points. These are differential geometrically, complex algebraic surfaces with positive curvatures, although the existence of K\"ahler-Einstein metric for such surfaces is an open problem.

From the Riemann-Roch theorem for surface (Noether's formula)
\begin{eqnarray}
12 (1 + p_a) & = & K_X^2 + c_2, \nonumber
\end{eqnarray}
where $p_a$ is the arithmetic genus of the surface $X$, which is known to be equal to $0$ (Castelnuovo) for rational surfaces. As the del Pezzo surfaces are the special cases of rational surfaces, the left hand side is $12$.
The second Chern class $c_2$ is
\begin{eqnarray}
c_2 & = & 12 - K_X^2 = 3 + n. \nonumber
\end{eqnarray}
Now that the second Chern character is, keeping in mind that $c_1 (X) = - K_X$,
\begin{eqnarray}
ch_2 (X) & = & - c_1 (X)^2 + 2 c_2 (X) \nonumber\\
& = & - (9 - n) + 2 (3 + n) \nonumber\\
& = & -3 + 3n. \label{RiemannRoch}
\end{eqnarray}
I conclude that the second Chern character vanishes if and only if $n = 1$. I will expect that the anomaly $2$-form (see section $3.2$) of Nekrasov\cite{N} for $1$ point blowup should be cancelled by the ``antisymmetric $\mu$-term" introduced by Nekrasov that is examined in this paper in more details, which will be defined in section $2.4$.

The generalization to non-toric surfaces and thereby to the ``re-classification by quantum invariants" of compact algebraic / complex surfaces in terms of deformed chiral algebras is interesting, but I did not pay attention to such cases. It was because I put the emphasis on the comparison or ``duality" of geometric models of different definitions. Thus one obtains a ``mixed anomaly cancellation" phenomenon at the case of $n = 1$; that is, $1$-point blowup of $\mathbb{CP}^2$. One did not explain the birationality of the obstraction class of gerbes of chiral de Rham complex, either.
\subsection*{Acknowledgments}
Without the communications to any individuals, this letter could not be published.

I would like to thank to the support and long-term discussion with Tomohide Terasoma. I cordially thank to Yosuke Imamura for his verification of my computations on the reviews and extensions of Malikov-Schechtman-Vaintrob\cite{MSV}, Nekrasov\cite{N}, and Witten\cite{Witten}. I also thank Edward Frenkel for his comment on Frenkel-Losev\cite{FL} and Malikov-Schechtman\cite{MS2}. I am indebted to the great works and the travel support of Alexander Beilinson.

The work of the author was in part done internationally in the Spring School on Superstring Theory and Related Topics, ICTP Italy (2004). Then, it was at the Strings 2005 annual meeting, (poster) presentation, the Fields Institute, Toronto in Canada (2005).

It was also at the AMS Summer Institute on Algebraic Geometry, Seattle in the U.S.A. (2005). It was also at the Affine Hecke Algebras, the Langlands Program, and Conformal Field Theory and Matrix Models held by CIRM in France (2006), the APS-JPS joint meeting, (an oral) presentation, Hawaii in the U.S.A. (2006), and the Geometric Langlands Program held by Lorentz Center Leiden (2008).

The partial financial support was from both the GCOE Tokyo and the Duisburg-Essen University for the Conference on Algebraic Geometry and Arithmetic, Germany (2010). The travel support was provided for the April visit to the University of Chicago, in the U.S.A. (2014), with the academic and financial support of Alexander Beilinson.
\section{OPE of chiral de Rham complex:\\Malikov-Schechtman-Vaintrob and Nekrasov}
In the following sections, the base field of ``$\mathcal{D}-$schemes" will be the complex ``number field" and the CFTs will be defined over a complex projective curves (compact Riemann surfaces), rather than higher dimensional CFTs. This ``number field" is neither a finite extension of rational numbers $\mathbb{Q}$ nor its algebraic closure $\bar{\mathbb{Q}}$ for the later consideration of existence of blowups. I will check the curved target space of $\beta \gamma$ CFT, which was examined by Witten\cite{Witten} in the case of $\mathbb{CP}^2$.
\subsection{Heisenberg algebra of bosons}
\footnote{As compared to the original paper of Malikov-Schechtman-Vaintrob, change the notation such that $b \mapsto \gamma, a \mapsto \beta$ with a minus factor to the original paper.}If one defines the commutation relation.
\newpage
\begin{eqnarray}
[(\beta_i)_n, \gamma^j_m]_- & = & \delta_i^j \delta_{m, -n} C, \nonumber\\
\gamma^i (z) & = & \sum_{n \in \mathbb{Z}} \gamma^i_n z^{-n}, \nonumber\\
\beta_i (z) & = & \sum_{n \in \mathbb{Z}} (\beta_i)_n z^{-n-1}, \nonumber
\end{eqnarray}
where $C$ is a constant that will be mapped to $1$ later in the polynomial ring. $\gamma$ has conformal dimension $0$, and $\beta$ has $1$. Namely, $\gamma^i_n (n \ge 1)$ are the annihilation operators, and $\gamma^i_n (n \le 0)$ are the creation operators. And $(\beta_i)_n (n \ge 0)$ are the annihilation operators, and $(\beta_i)_n (n \le -1)$ are the creation operators.

I will regard $\gamma^i$ as the coordinate, and $\beta_i$ are vector fields. For $x = (\beta_i)_n$, or $\gamma^i_n (n \in \mathbb{Z})$ and $B \in End(V_N)$ ($V_N$ is the state space that the Heisenberg algebras act on, which was defined after the (Fock) vacuum of highest weight state), the normal ordered product $:xB:$ is given by
\begin{eqnarray}
:xB: = \left\{ \begin{array}{ll}
Bx & \mbox{(if $x$  is an annihilation operator),} \\
xB & \mbox{(otherwise).} \nonumber
\end{array} \right.
\end{eqnarray}

If it is assumed $|z| > |w|$, which is called the radial ordering, in order to make the summation absolutely convergent, one obtains the following OPE.
\begin{eqnarray}
\beta_i (z) \gamma^j (w) & = & \frac{\delta_i^j}{z-w}+(regular). \nonumber
\end{eqnarray}
Similarly,
\begin{eqnarray}
\beta_i (z) \beta_j (w) & = & (regular), \nonumber\\
\gamma^i (z) \gamma^j (w) & = & (regular). \nonumber
\end{eqnarray}

The stress energy tensor is given by
\begin{eqnarray}
L(z) & = & :\partial_z \gamma (z) \beta(z):. \nonumber
\end{eqnarray}
OPE of the stress-energy tensor is given by
\begin{eqnarray}
L(z) L(w) & \sim & \frac{1}{(z-w)^4} + \frac{2 L (w)}{(z-w)^2} + \frac{\partial_w L(w)}{z-w}. \nonumber
\end{eqnarray}
This can be verified by using the Wick theorem. $\sim$ is the OPE, which ignores regular terms.

\subsection{Clifford algebra of chiral fermions}

\footnote{As compared to the original paper of Malikov-Schechtman-Vaintrob, I change the notation such that $\phi \mapsto b, \psi \mapsto c$.}Now I will consider the fermion fields $b_i, c^j$. The conformal dimension of $b_i$ and $c^j$ are 0 and 1, respectively. OPEs of $bc$-$\beta\gamma$ system are given by
\begin{eqnarray}
\beta_i (z) \gamma^j (w) & = & \frac{\delta_i^j}{z - w}, \label{betagamma_OPE} \\
\gamma^i(z) \gamma^j (w) & \sim & 0, \nonumber\\
\beta_i (z) \beta_j (w) & \sim & 0, \nonumber\\
b^i (z) c_j (w) & = & \frac{\delta^i_j}{z - w}, \nonumber\\
b^i (z) b^j (w) & \sim & 0, \nonumber\\
c_i (z) c_j (w) & \sim & 0, \nonumber\\
\gamma^i (z) b^j (w) & \sim & 0, \nonumber\\
\gamma^i (z) c_j (w) & \sim & 0, \nonumber\\
\beta_i (z) b^j (w) & \sim & 0, \nonumber\\
\beta_i (z) c_j (w) & \sim & 0. \nonumber
\end{eqnarray}
\subsection{A topological vertex algebra of rank $D$}

In order to describe the topological vertex algebra, I introduce the following fields $L$: stress-energy tensor of conformal dimension $2$, $J$: $U(1)$ current of conformal dimension $1$, and $Q, G$ are (twisted version of) two generators of $\mathcal{N}=2$ supersymmetry of conformal dimension $1$ and $2$, respectively. They are written as follows:
\begin{eqnarray}
L & = & \sum_i \left[ :\partial \gamma^i (z) \beta^i (z): + :\partial b^i (z) c^i (z): \right], \nonumber\\
J & = & \sum_i :b^i (z) c^i (z):, \nonumber\\
Q & = & \sum_i :\beta^i (z) b^i (z):, \nonumber\\
G & = & \sum_i :c^i(z) \partial \gamma^i (z):. \nonumber
\end{eqnarray}
\newpage
The OPEs are given by
\begin{eqnarray}
L(z) L(w) & = & \frac{2 L(w)}{(z-w)^2} + \frac{\partial_w L(w)}{z-w}, \nonumber\\
J(z) J(w) & = & \frac{D}{(z-w)^2}, \nonumber\\
L(z) J(w) & = & - \frac{D}{(z-w)^3} + \frac{J(w)}{(z-w)^2} + \frac{\partial_wJ(w)}{z-w}, \nonumber
\end{eqnarray}
where no special consideration for supersymmetry algebras is required. However, the $2$ supersymmetry generators arise as follows:
\begin{eqnarray}
G(z) G(w) & = & 0, \nonumber\\
L(z) G(w) & = & \frac{2 G(w)}{(z-w)^2} + \frac{\partial_w G(w)}{z-w}, \nonumber\\
J(z) G(w) & = & - \frac{G(w)}{z-w}, \nonumber\\
Q(z) Q(w) & = & 0, \nonumber\\
L(z) Q(w) & = & \frac{Q(w)}{(z-w)^2} + \frac{\partial_w Q(w)}{z-w}, \nonumber\\
J(z) Q(w) & = & \frac{Q(w)}{z-w}, \nonumber\\
Q(z) G(w) & = & \frac{D}{(z-w)^3} + \frac{J(w)}{(z-w)^2} + \frac{L(w)}{z-w}. \label{QG_OPE}
\end{eqnarray}
Here $D$ denotes the complex dimension $\dim_{\mathbb{C}} X$ of the target manifolds. Note that the operator $L(z)$ can be expressed as the commutator. This is derived by integrating the equation (\ref{QG_OPE}) around $w$ and one obtains
\begin{eqnarray}
[Q_0, G(w)]_+ & = & L(w) \nonumber
\end{eqnarray}
where $\displaystyle{Q_0 = \oint Q (z) dz}$ is the residue after contour integration around the origin of
$Q(z) = \sum_{n \in \mathbb{Z}} Q_n z^{n-1}$.
Since the stress tensor $L(z)$ becomes BRST (cohomology) exact, one obtains a topological theory. This algebra is determined by the twist of $\mathcal{N}=2$ SCFT (Super Conformal Field Theory), whose stress energy tensor is $\displaystyle{T=L - \frac12 \partial J}$. 

\subsection{Useful OPEs}

The purpose of this section is to derive useful formulae which will be used in the next section.
In the next section I want to calculate the OPE for
\begin{eqnarray}
\tilde{\beta}_a & := & \beta_i g^i_a + B_{ai} \partial \gamma^i \nonumber\\
& = & J_{g_a} + C_{B_a}, \nonumber
\end{eqnarray}
where the $g^i_a$ and $B_{ai}$ are the functions of $\gamma$ and I will abbreviate summations over indices when no confusion occurs. This is the notation of Einstein in the tensor calculus. I define $J_V$ and $C_B$ for every $1$-form $B \in \Omega^1_U$, and every vector $V \in T_U$, where $U$ is the coordinate patch which one is working with
\begin{eqnarray}
J_V & = & : \beta_i V^i (\gamma) (z) : \nonumber\\
& := & \lim_{\epsilon \to 0} [ \beta_i (z + \epsilon) V^i (\gamma (z)) - \frac{1}{\epsilon} \partial_i V^i (\gamma (z))], \nonumber\\
C_B & = & B_i (\gamma (z)) \partial \gamma^i. \nonumber
\end{eqnarray}
Notice that $C_B$ has the conformal weight $1$, whereas $J_V$ has also conformal dimension $1$ but has an extra $(z-w)^{-3}$ term in the OPE with the stress energy tensor $L(z)$.

By utilizing the inner product, the Lie derivatives, and the commutation relation in the basis of tangent bundle $T_U$, one obtains
\begin{eqnarray}
[V_a, V_b]^j (z) & = & \partial_i V^j_b (z) V^i_a (z) - \partial_i V^j_a (z) V^i_b (z), \nonumber\\
\mathcal{L}_V B (z) & = & \partial_i B_j (z) V^i (z) + \partial_j V^i (z) B_i (z). \nonumber
\end{eqnarray}
We can compute the following OPEs
\begin{eqnarray}
\quad J_{V_a} (z + \epsilon) J_{V_b} (z) & \sim & - \frac{\Sigma_{ab} (z + \epsilon) + \Sigma_{ab} (z)}{2\epsilon^2} + \frac{J_{[ V_a, V_b] (z)}}{\epsilon} - \frac{C_{\Omega_{ab}}(z)}{\epsilon}, \label{JJ_OPE} \\
\quad J_V (z + \epsilon) C_B (z) & \sim & \frac{\iota_V B(z)}{\epsilon^2} + \frac{C_{\mathcal{L}_V B (z)}}{\epsilon}, \label{JC_OPE} \\
C_B (z + \epsilon) C_{B'} (z) & \sim & 0. \nonumber
\end{eqnarray}
Here
\begin{eqnarray}
\Sigma_{ab} & = & tr \mathcal{V}_a \mathcal{V}_b, \nonumber\\
\Omega_{ab} & = & \frac12 tr(\mathcal{V}_a d \mathcal{V}_b - \mathcal{V}_b d \mathcal{V}_a), \nonumber
\end{eqnarray}
where the matrix is defined by
\begin{eqnarray}
(\mathcal{V}_a)_{ij} & = & \partial_i V^j_a. \nonumber
\end{eqnarray}
I will use these equations in the following sections.

\subsection{OPE on generalized complex manifolds}

For later use, I will compute the OPE of combination of tangent and cotangent (or normal) bundle. $V, W \in T_X$ and $\xi, \eta \in \Omega^1_X$
\begin{eqnarray}
v & = & V \oplus \xi, \quad w = W \oplus \eta \in T_X \oplus \Omega^1_X, \nonumber\\
\mathcal{O}_v & := & J_V + C_{\xi} \quad \mathcal{O}_w := J_W + C_{\eta}. \nonumber
\end{eqnarray}
Using the formula of the last section
\begin{eqnarray}
\mathcal{O}_v (z + \epsilon) \mathcal{O}_w (z) & \sim &  J_V (z + \epsilon) J_W (z) + C_{\xi} (z + \epsilon) J_W (z) \nonumber\\
& &  + J_V (z + \epsilon) C_{\eta} (z). \nonumber
\end{eqnarray}
Then I use (\ref{JJ_OPE}), (\ref{JC_OPE}) and obtain
\begin{eqnarray}
LHS & \sim & \left[ - \frac{1}{2 \epsilon^2} (\Sigma_{VW} (z + \epsilon) + \Sigma_{VW} (z)) + \frac{1}{\epsilon} J_{[V,W]} (z) - \frac{1}{\epsilon} C_{\Omega_{VW}} (z) \right] \nonumber\\
& + & \left[ \frac{1}{(- \epsilon)^2} \iota_W \xi (z + \epsilon) + \frac{1}{(-\epsilon)} C_{\mathcal{L}_W \xi} (z + \epsilon) \right] \nonumber\\
& + & \left[ \frac{1}{\epsilon^2} \iota_V \eta (z) + \frac{1}{\epsilon} C_{\mathcal{L}_V \eta} (z) \right]. \label{LHS}
\end{eqnarray}
By defining the metric in the generalized complex manifolds as follows:
\begin{eqnarray}
2 g (v, w) & := & -\Sigma_{VW} + \iota_V \eta + \iota_W \xi, \nonumber
\end{eqnarray}
we can rewrite the $O(1/\epsilon^2)$ part of (\ref{LHS})
\begin{eqnarray}
\frac{1}{\epsilon^2} (g(v,w) (z + \epsilon) + g(v,w) (z)) + \frac{1}{\epsilon} (\mathcal{O}_{[[v,w]]} - C_{\Omega_{VW}}), \label{LHS2}
\end{eqnarray}
where I defined the Courant bracket
\begin{eqnarray}
[[v,w]] & := & [V,W] + \mathcal{L}_V \eta - \mathcal{L}_W \xi - \frac12 d (\iota_V \eta - \iota_W \xi). \nonumber
\end{eqnarray}

\subsection{The target space coordinate transformations}
Let us introduce an Ansatz of Nekrasov(see \cite{N}) for a coordinate change $\gamma^i$ to $\tilde{\gamma}^a$ on the condition that the OPE of $\beta \gamma$-system is unchanged after coordinate transformations. I write the Ansatz as follows
\begin{eqnarray}
\tilde{\beta}_a & := & \beta_i g^i_a + B_{ai} \partial \gamma^i \nonumber\\
& = & J_{g_a} + C_{B_a}, \label{NekrasovAnsatz}
\end{eqnarray}
where $B_a \in \Omega^1_U, g_a \in T_U$. Then by requiring
\begin{eqnarray}
\tilde{\beta}_a (z) \tilde{\gamma}^b (w) & \sim & \frac{\delta^b_a}{z - w}, \nonumber
\end{eqnarray}
we obtain
\begin{eqnarray}
g^i_a & := & \frac{\partial \gamma^i}{\partial \tilde{\gamma}^a}. \nonumber
\end{eqnarray}
Now I examine the OPE relation
\begin{eqnarray}
\tilde{\beta}_a \tilde{\beta}_b & \sim & 0. \label{betabetaOPE}
\end{eqnarray}
By using the OPE of (\ref{LHS2}) of generalized complex manifolds of $v =(g_a \oplus B_a), w = (g_b \oplus B_b) \in T_U \oplus \Omega^1_U$, one obtains
\begin{eqnarray}
\tilde{\beta}_a (z + \epsilon) \tilde{\beta}_b (z) & \sim & \frac{1}{\epsilon^2} (g(v,w) (z + \epsilon) + g(v,w) (z)) \nonumber\\
& + & \frac{1}{\epsilon} (\mathcal{O}_{[[v,w]]} - C_{\Omega_{ab}}). \label{beta_beta}
\end{eqnarray}
By requiring (\ref{betabetaOPE}), the $O(1/\epsilon)$ and $O(1/\epsilon^2)$ terms have to vanish. As for $O(1/\epsilon^2)$ part, one has
\begin{eqnarray}
g(v,w) (z) & = & - \Sigma_{ab} + \iota_{g_a} B_b + \iota_{g_b} B_a \nonumber\\
& = & 0. \label{sigma_B}
\end{eqnarray}
We now would like to determine $B_{a}$. Let us define the symmetric $\sigma_{ab}$ and antisymmetric $\mu_{ab}$ part of $B_a$ as follows
\begin{eqnarray}
B_a & = & \frac12 (\sigma_{ab} - \mu_{ab}) d \tilde{\gamma}^b. \nonumber
\end{eqnarray}
Then, from the definition of symmetric part and (\ref{sigma_B}), one can first conclude that
\begin{eqnarray}
\sigma_{ab} = \sigma_{ba} & = & \iota_{g_a} B_b + \iota_{g_b} B_a \nonumber\\
& = & \Sigma_{ab} = \sum_{i, j} \partial_i g^j_a \partial_j g^i_b. \nonumber
\end{eqnarray}
We still have to determine $\mu_{ab}$. From the $O(1/\epsilon)$ part of $\tilde{\beta} \tilde{\beta}$ OPE (\ref{beta_beta}), one obtains the necessary equation
\begin{eqnarray}
\mathcal{O}_{[[v,w]]} - C_{\Omega_{ab}} & = & J_{[g_a, g_b]} + C_{\mathcal{L}_{g_a} B_b - \mathcal{L}_{g_b} B_a - \frac12 d \mu_{ab}} - C_{\Omega_{ab}} \nonumber\\
& = &  C_{\mathcal{L}_{g_a} B_b - \mathcal{L}_{g_b} B_a - \frac12 d \mu_{ab}} - C_{\Omega_{ab}} \nonumber\\
& = & 0. \nonumber
\end{eqnarray}
I contract this equation with any vector $g_c \in T_U$ to obtain the equation
\begin{eqnarray}
\mathcal{L}_{g_c} \mu_{ab} + 2 \iota_{g_c} \iota_{g_a} d B_b- 2 \iota_{g_c} \iota_{g_b} d B_a & = & tr (\mathcal{G}_a \mathcal{L}_{g_c} \mathcal{G}_b - \mathcal{G}_b \mathcal{L}_{g_c} \mathcal{G}_a), \nonumber
\end{eqnarray}
where $\mathcal{G}_a$ is defined by
\begin{eqnarray}
(\mathcal{G}_a)_{ij} = \partial_i g_a^j. \nonumber
\end{eqnarray}
By using the Maurer-Cartan equation for $\mathcal{G}$, this equation can be rewritten as
\begin{eqnarray}
d \mu & = & - tr (d \tilde{g}^a_{\ i} (\tilde{g}^{-1})^i_{\ b})^3 \nonumber\\
& = & tr (g^{-1} d g)^3, \nonumber
\end{eqnarray}
where $\tilde{g}^b_{\ i} g^i_{\ a} = \delta^b_{\ a}$ and there are no further conditions for $\mu$ besides this equation.
\section{Witten's $\mathbb{CP}^2$ case}
The ``toric" diagrams of target manifolds will be carefully considered at the next section of blowups. In this section, for your convenience and reference of curiosity, let us just write down how this toric diagram is also drawn where no exceptional divisor is necessary. I do not utilize this toric diagram directly in this section.
\begin{figure}[htb]
\begin{center}
\includegraphics[width=6cm]{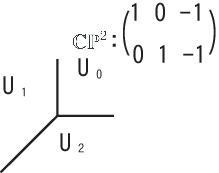}
\end{center}
\caption{Toric diagram for $\mathbb{CP}^2$}
\label{CP2}
\end{figure}
\subsection{Witten's $p_1$ anomaly for $\mathbb{CP}^2$}
Let us consider $\mathbb{CP}^2$ as the target space. I will review the computation of the $p_1$ anomaly of Witten\cite{Witten}. Let $U_{\alpha} \subset \mathbb{CP}^2$ be the affine coordinate patch defined by $\lambda_{\alpha} \neq 0$ ($\alpha = 0, 1, 2$) for projective coordinate $(\lambda_0: \lambda_1: \lambda_2)$, Then, for the coordinate $\gamma^i$ ($i=1, 2$) of patch $U_{\alpha}$, one has the OPE of (\ref{betagamma_OPE})
\begin{eqnarray}
\beta_i (z) \gamma^j (z') \sim + \frac{\delta_i^j}{z - z'}. \nonumber
\end{eqnarray}
Under the combined coordinate transformations from $U_0$ to $U_0$: $U_0 \to U_1 \to U_2 \to U_0$
\begin{eqnarray}
\gamma^j \to \gamma^j, \nonumber\\
\beta_i \to \beta_i' = \beta_i + f_{ij} \partial \gamma^j, \label{WittenAnsatz}
\end{eqnarray}
where $f_{ij} = - f_{ji}$ is an antisymmetric tensor.\\
This is the Ansatz of Witten\cite{Witten}, which was originally derived only from the pole analysis and does not have a consideration by the antisymmetric $\mu$-term of Nekrasov\cite{N}.
I will denote $\gamma^1, \gamma^2$ as $v$, $w$ and $\beta_1, \beta_2$ as $V$, $W$. I assume that the generator of $H^2(\mathbb{CP}^2, \Omega^{2,cl})$ is given by
\begin{eqnarray}
\frac12 f_{ij} (\gamma) d \gamma^i \wedge d \gamma^j = \frac{dv \wedge dw}{vw}, \nonumber
\end{eqnarray}
where $v$, $w$ are the inhomogeneous standard coordinate of $U_0 \subset \mathbb{CP}^2$:
\begin{eqnarray}
v & = & \frac{\lambda_1}{\lambda_0}, \nonumber\\
w & = & \frac{\lambda_2}{\lambda_0}. \nonumber
\end{eqnarray}
$v$ and $w$ are symbols more fundamental than $v^{[i]}$ and $w^{[i]}$, ($i = 1, 2, 3$ mod $3$), which will be defined for the coordinate transformations of the following subsections. Then I perform successive coordinate transformations $U_0 \to U_1 \to U_2$ and back to $U_0$. I first check the coordinate transformation of $U_0 \to U_1$, which can be read from the toric diagram (Figure \ref{CP2}).

\subsubsection{\underline{$U_0 \to U_1$}}

In the following subsections, as Witten\cite{Witten} defined, I will use the notation $v^{[i]}, w^{[i]}, V^{[i]}, W^{[i]}$, instead of $\gamma, \beta$. To be more precise, let

\begin{eqnarray}
v^{[i]} & = & \frac{\lambda_{i+1}}{\lambda_i}, \nonumber\\
w^{[i]} & = & \frac{\lambda_{i+2}}{\lambda_i}. \nonumber
\end{eqnarray}

$V^{[i]}, W^{[i]}$ are the corresponding $1$-form.

\begin{eqnarray}
v^{[1]} & = & \frac{w}{v}, \nonumber\\
w^{[1]} & = & \frac{1}{v}, \nonumber\\
V^{[1]} & = & v W, \label{V1}\\
W^{[1]} & = & -v^2 V - vw W - \frac{5}{2} \label{W1} \partial v. \nonumber
\end{eqnarray}
The Jacobian matrix $g_i^a$ is given by
\begin{eqnarray}
g_i^a = (g)_{ia} & = & \frac{\partial \tilde{\gamma}^a}{\partial \gamma^i} \nonumber\\
& = & \left( \begin{array}{cc} - w / v^2 & - 1 / v^2 \\ 1 / v & 0 \end{array} \right)_{ia}. \nonumber
\end{eqnarray}
Then the inverse matrix is
\begin{eqnarray}
g^i_a = (g^{-1})_{ai} & = & \left( \begin{array}{cc} 0 & v \\ - v^2 & -vw \end{array} \right)_{ai}. \nonumber
\end{eqnarray}
This gives the Jacobian part $J_{g_a}$ of \textbf{Nekrasov's formula} (\ref{NekrasovAnsatz}) that assumes the generalized complex structure. Now that I have to determine the $B_{ai}$ of
\begin{eqnarray}
\tilde{\beta_a} & := & \beta_i g_a^i + B_{ai} \partial \gamma^i \nonumber \\
& = & J_{g_a} + C_{B_a}. \nonumber
\end{eqnarray}
I decompose $B$ as $B_a = \displaystyle{\frac12 (\sigma_{ab} - \mu_{ab}) d \tilde{\gamma}^b}$, where $d \mu = tr (g^{-1} dg)^3 = 0$ since $\mathbb{CP}^2$ is $2$ dimensional. On the other hand, the symmetric part is given by
\begin{eqnarray}
\sigma_{ab} & = & \partial_i g_a^j \partial_j g^i_b. \nonumber
\end{eqnarray}
Then,
\begin{eqnarray}
(B)_{ia} = B_{ai} & = & \frac12 g^b_i \sigma_{ab} \nonumber\\
& = & \frac12 \left( \begin{array}{cc} 0 & -5 \\ 0 & 0 \end{array} \right)_{ia}. \nonumber
\end{eqnarray}

\subsubsection{\underline{$U_0 \to U_1 \to U_2$}}

Similar discussions yield that the coordinate\\
transformations for $U_1 \to U_2$
\begin{eqnarray}
v^{[2]} & = & \frac{w^{[1]}}{v^{[1]}} = \frac{1}{w}, \nonumber\\
w^{[2]} & = & \frac{1}{v^{[1]}} = \frac{v}{w}, \nonumber\\
V^{[2]} & = & v^{[1]} W^{[1]}, \label{V2} \nonumber\\
W^{[2]} & = & -(v^{[1]})^2 V^{[1]} - v^{[1]} w^{[1]} W^{[1]} - \frac{5}{2} \partial v^{[1]}, \label{W2}
\end{eqnarray}
and $U_2 \to U_0$
\begin{eqnarray}
v^{[3]} & = & \frac{w^{[2]}}{v^{[2]}} = v, \nonumber\\
w^{[3]} & = & \frac{1}{v^{[2]}} = w, \nonumber\\
V^{[3]} & = & v^{[2]} W^{[2]}, \nonumber\\
W^{[3]} & = & -(v^{[2]})^2 V^{[2]} - v^{[2]} w^{[2]} W^{[2]} - \frac{5}{2} \partial v^{[2]}. \nonumber
\end{eqnarray}
\newpage
To be more careful, when one substitutes and combines the equations\\
(\ref{V1})(\ref{W1}) ($U_0 \to U_1$) in the equations of $V^{[2]}$ (\ref{V2}) and $W^{[2]}$ (\ref{W2}) for $U_1 \to U_2$, which have the cross terms with $\displaystyle{\frac{1}{v}}$ or $\displaystyle{\frac{w}{v}}$.
Since the definition of the normal product depends on the patches and each term of (\ref{W2}) is defined by the normal product, the first term of $W^{[2]}$, for example, is defined as
\begingroup\makeatletter\def\f@size{10}\check@mathfonts
\begin{eqnarray}
-(v^{[1]})^2 V^{[1]} & = & \lim_{z' \to z} (-v^{[1]} (z')^2 V^{[1]}(z) + 2 v^{[1]}(z') \frac{1}{z' - z}) \nonumber\\
& = & \lim_{z' \to z} \left[ - (\frac{w(z')}{v(z')})^2 v(z) W (z) + 2 (\frac{w(z')}{v(z')}) \frac{-1}{z' - z}) \right] \nonumber\\
& = & \lim_{z' \to z} \left[ - \frac{2 w (z')}{z - z'} \frac{v(z)}{v(z')^2} - (\frac{w(z')}{v(z')})^2 v(z) W(z) + 2 (\frac{w(z')}{v(z')}) \frac{-1}{z' - z} \right] \nonumber\\
& = & \lim_{\epsilon \to 0} \left[ (\frac{2}{\epsilon} \frac{w}{v} - 4 \frac{w \partial v}{v^2} + \frac{2 \partial w}{v}) - \frac{w^2}{v}W + (\frac{2(\partial v w - \partial w v)}{v^2} - \frac{2}{\epsilon}\frac{w}{v}) \right] \nonumber\\
& = & -2 \frac{\partial v w}{v^2} - \frac{w^2}{v} W. \nonumber
\end{eqnarray}
\endgroup
By applying similar treatments for the second term of (\ref{W2}), I obtained the relation for $U_0 \to U_1 \to U_2$. Therefore
\begin{eqnarray}
V^{[2]} & = & -vwV -w^2 W - \frac{3 w \partial v}{2v} - \partial w, \nonumber\\
W^{[2]} & = & wV - \frac{3 \partial w}{2v}. \nonumber
\end{eqnarray}

\subsubsection{\underline{$\displaystyle{U_0 \to U_1 \to U_2 \to U_0}$}}
Once more, one will substitute the $\displaystyle{V^{[3]}}$, $\displaystyle{W^{[3]}}$ by the equations above to obtain
\begin{eqnarray}
V^{[3]} & = & v^{[2]} W^{[2]} \nonumber\\
& = & \frac{1}{w} :w V - \frac{3}{2} \frac{\partial w}{v}: \nonumber\\
& = & V - \frac{3}{2} \frac{\partial w}{v w} \nonumber,
\end{eqnarray}
\newpage
and for $z'=z + \epsilon$
\begingroup\makeatletter\def\f@size{10}\check@mathfonts
\begin{eqnarray}
W^{[3]} & = & - :(v^{[2]})^2 V^{[2]}: - :v^{[2]} w^{[2]} W^{[2]}: -\frac{5}{2} \partial v^{[2]}\nonumber\\
& = & - \lim_{\epsilon \to 0} \left[ (v^{[2]} (z'))^2 V^{[2]} (z) + 2 V^{[2]} (z) \frac{1}{\epsilon} + 2 \partial v^{[2]} \right] \nonumber\\
& & - \lim_{\epsilon \to 0} \left[ v^{[2]} (z') w^{[2]} (z') W^{[2]} (z) + v^{[2]} \frac{1}{\epsilon} + \partial v^{[2]} \right] - \frac{5}{2} \partial v^{[2]} \nonumber\\
& = & \lim_{\epsilon \to 0} \left[ - \frac{3}{\epsilon} v^{[2]} (z) - \frac{11}{2} \partial v^{[2]} - (v^{[2]} (z'))^2 V^{[2]} (z) - v^{[2]} (z') w^{[2]} (z') W^{[2]} (z) \right] \nonumber\\
& = & \lim_{\epsilon \to 0} \left[ -\frac{5}{2} \partial v^{[2]} + \frac{1}{w(z')^2} (vwV + w^2W + \frac{3 w \partial v}{2v} + \partial w) (z) \right] \nonumber\\
& & + \lim_{\epsilon \to 0} \left[ - \frac{1}{w(z')}\frac{v(z')}{w(z')} (w V - \frac{3}{2} \frac{\partial w}{v}) (z) \right] \nonumber\\
& = & \lim_{\epsilon \to 0} \left[ - \frac{3}{\epsilon} v^{[2]} (z) - \frac{11}{2} \partial v^{[2]} + \frac{v V}{w} + \frac{2 w^{-3} (z') w^2(z)}{\epsilon} + \frac{3 \partial v}{2 v w} \right] \nonumber\\
& & + \lim_{\epsilon \to 0} \left[ \frac{\partial w}{w^2} + \frac{w (z)}{w(z')^2} \frac{1}{\epsilon} - \frac{v V}{w} + \frac{3 \partial w}{2 w^2} + W(z) \right] \nonumber\\
& = & W(z) + \frac{3}{2} \frac{\partial v}{vw}. \nonumber
\end{eqnarray}
\endgroup
Thereby, in total,
\begin{eqnarray}
v \to v ,\nonumber\\
w \to w ,\nonumber\\
V \to V - \frac{3}{2} \frac{\partial w}{vw}, \nonumber\\
W \to W + \frac{3}{2} \frac{\partial v}{vw}. \label{VWcyclic}
\end{eqnarray}
This is certainly the Ansatz of Witten (\ref{WittenAnsatz}). I will check whether this result can be derived from the Ansatz of Nekrasov (\ref{NekrasovAnsatz}) with $d \mu = 0$ whereas the total gerbe term $f_{ij}$ is antisymmetric. This is due to the anomaly $2$-form of Nekrasov.

\subsection{Anomaly $2$-form $\psi_{\alpha \beta \gamma}$ after Nekrasov's lecture}

For $U_{\alpha}, U_{\beta},$\\
$ U_{\gamma}$: affine patches, I perform successive coordinate transformations\\
$U_{\alpha} \to U_{\beta} \to U_{\gamma} \to U_{\alpha}$.
The coordinate transformations in each step are given by
\begin{eqnarray}
U_{\alpha} \to U_{\beta} & : & \gamma^i \to \gamma^a, \quad
\beta^i \to \beta^a = g^a_i \beta^i + B_{ai} \gamma^i, \nonumber\\
U_{\beta} \to U_{\gamma} & : &\gamma^a \to \gamma^p, \quad
\beta^a \to \beta^p = g^p_a \beta^a + B'_{p a} \gamma^a, \nonumber\\
U_{\gamma} \to U_{\alpha} & : &\gamma^p \to \gamma^I, \quad
\beta^p \to \beta^I = g^I_p \beta^p + B''_{I p} \gamma^p.\label{3rdstep}
\end{eqnarray}
I will use the indices $i,j,k,\cdots,a,b,c,\cdots,p,q,r,\cdots,I,J,K$ as
\begin{eqnarray}
\gamma^i, \gamma^j, \gamma^k \in U_{\alpha}, \nonumber\\
\gamma^a, \gamma^b, \gamma^c \in U_{\beta}, \nonumber\\
\gamma^p, \gamma^q, \gamma^r \in U_{\gamma}, \nonumber\\
\gamma^I, \gamma^J, \gamma^K \in U_{\alpha}. \nonumber
\end{eqnarray}
By combining these three coordinate transformations, I obtain
\begin{eqnarray}
\gamma^i & \to & \gamma^i, \nonumber\\
\beta_j & \to & \beta_j -\frac{1}{2} (\psi_{\alpha \beta \gamma})_{Ij} \partial \gamma^I. \label{Btot}
\end{eqnarray}
$\psi_{\alpha \beta \gamma}$ is defined by
\begin{eqnarray}
\psi_{\alpha \beta \gamma} & = & \mu_{\alpha \beta} + \mu_{\beta \gamma} + \mu_{\gamma \alpha} - tr(g'' dg' \wedge dg) \nonumber\\
& =: & \mu_{\alpha \beta} + \mu_{\beta \gamma} + \mu_{\gamma \alpha} - \psi^0_{\alpha \beta \gamma} \nonumber\\
& & \in H^2(X, T_X \oplus \Omega^2_X / d \Omega^1_X) \label{Psi},
\end{eqnarray}
where $\mu_{\alpha \beta}$, $\mu_{\beta \gamma}$, and $\mu_{\gamma \alpha}$ are the antisymmetric parts of $B$, $B'$, and $B''$. This is called the anomaly $2$-form of Nekrasov\cite{N}.
[$T_X$ is for the gravitational anomaly coming from the product of Weyl anomaly and gravitational anomaliy from $c_1$ (world-sheet) $\times$ $c_1$ (target space) and I do not use nor explain it.]
 (\ref{Btot}) can be proved as follows: for $2$ step coordinate changes $U_{\alpha} \to U_{\beta} \to U_{\gamma}$,
\begingroup\makeatletter\def\f@size{10}\check@mathfonts
\begin{eqnarray}
\beta'_p & = & g^a_p \tilde{\beta}_a + B'_{p a} \partial \gamma^a \nonumber\\
& = & \lim_{\epsilon \to 0} \left[ g^a_p (z + \epsilon) \tilde{\beta}_a (z) + \frac{1}{\epsilon} \partial_a g^a_p (z + \epsilon) \right] \nonumber\\
& = & \lim_{\epsilon \to 0} \left[ g^a_p (z + \epsilon) (g^i_a (z) \beta_i (z) + B_{ai} (z) \partial \gamma^i (z)) + \frac{1}{\epsilon} \partial_a g^a_p (z + \epsilon) \right] \nonumber\\
& = & \lim_{\epsilon \to 0} \left[ g^a_p g^i_a \beta_i + g^a_p B_{ai} (z) \partial \gamma^i (z) - \frac{1}{\epsilon} \partial_i g^a_p (z + \epsilon) g^i_a (z) + \frac{1}{\epsilon} \partial_a g^a_p (z + \epsilon) \right].\nonumber
\end{eqnarray}
\endgroup
This can be rewritten as follows by using $g^a_p g^i_a = g^i_p, g^i_a (z) \sim g^i_a (z + \epsilon) - \partial_z g^i_a \epsilon$ and $\partial_i g^a_{\alpha} (z + \epsilon) g^i_a (z + \epsilon) = \partial_a g^a_{\alpha} (z + \epsilon)$,
\begin{eqnarray}
\beta'_p & = & g^i_p \beta_i + g^a_p B_{ai} \partial \gamma^i + B'_{p a} g^a_i \partial \gamma^i + \partial_i g^a_p \partial_j g^i_p \partial \gamma^j \nonumber\\
& = & g^i_{\alpha} \beta_i + \hat{B}_{\alpha i} \partial \gamma^i, \label{2step}
\end{eqnarray}
where $\hat{B}_{\alpha i}$ is given by
\begin{eqnarray}
\hat{B}_{p i} & := & (B_{p i} + B'_{p i} + \partial_j g^a_p \partial_i g^j_a)  \partial \gamma^i. \nonumber
\end{eqnarray}
The third step coordinate change $U_{\alpha} \to U_{\beta} \to U_{\gamma} \to U_{\alpha}$ (\ref{Btot}) is obtained by combining (\ref{2step}) and  (\ref{3rdstep}).

Let us confirm that this will reproduce the previous result for the case of $\mathbb{CP}^2$, where $d \mu = 0$ and one can set $\mu$ as arbitrary value. One has to check whether $\psi_{\alpha \beta \gamma}=0$ for some $\mu$ after computing $\psi^0_{\alpha \beta \gamma}$
\begin{eqnarray}
(g^i_a)_{ai} & = & \left( \begin{array}{cc} 0 & v \\ - v^2 & -vw \end{array} \right), \nonumber\\
(g'^a_p)_{p a} & = & \left( \begin{array}{cc} 0 & v^{[1]} \\ - (v^{[1]})^2  & -v^{[1]} w^{[1]} \end{array} \right) \nonumber\\
& = & \left( \begin{array}{cc} 0 & w / v \\ - w^2 / v^2  & - w / v^2 \end{array} \right), \nonumber\\
(g''^p_I)_{I p} & = & \left( \begin{array}{cc} 0 & v^{[2]} \\ - (v^{[2]})^2  & -v^{[2]} w^{[2]} \end{array} \right) \nonumber\\
& = & \left( \begin{array}{cc} 0 & 1 / w \\ - 1 / w^2 & - v / w^2 \end{array} \right). \nonumber
\end{eqnarray}
Therefore
\begin{eqnarray}
dg' \wedge dg & = & dv \wedge dw \left( \begin{array}{cc} 2 & 2w/v \\ 2 / v & - w /v^2 \end{array} \right), \nonumber\\
tr g'' dg' \wedge dg & = & \frac{3 dv \wedge dw}{vw}. \nonumber
\end{eqnarray}
This agrees with the result of direct OPE computation in (\ref{VWcyclic}). And the anomaly exists because $\mu$ cannot eliminate this term. This result will be recalled as
\begin{eqnarray}
c_{123} & = & +3, \label{TotalAnomaly0}
\end{eqnarray}
in the chapters below. (The notation $c_{\alpha \beta \gamma}$ will be introduced in the definition (\ref{Psi}) of the partial anomaly $\displaystyle{\psi^0_{\alpha \beta \gamma}}$ and its coefficient $c_{\alpha \beta \gamma}$) This result will be referred as the total anomaly for $n=0$ point blowup of $\mathbb{CP}^2$.
\newpage
\section{Cases of del Pezzo surfaces}
\subsection{Anomaly $2$-forms for vanishing anomaly of $1$ point blowup of $\mathbb{CP}^2$}

\begin{figure}[htb]
\begin{center}
\includegraphics[width=6cm]{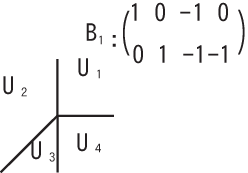}
\end{center}
\caption{Toric diagram for $1$ point blowup of $\mathbb{CP}^2$}
\label{Toric1}
\end{figure}

The coordinate change can be read from the toric diagram (Figure \ref{Toric1}) and (Figure \ref{Del1})

This schematic diagram of $4$ neighbourhoods of $1$ point blowup of $\mathbb{CP}^2$ can be rewritten as the following diagram:
\begin{figure}[htb]
\begin{center}
\includegraphics[width=9cm]{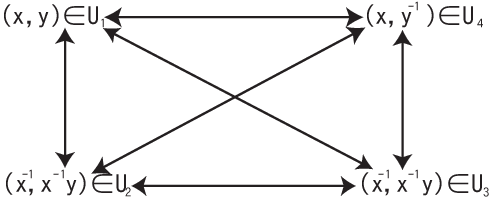}
\end{center}
\caption{Coordinate changes for $1$ point blowup of $\mathbb{CP}^2$}
\label{Del1}
\end{figure}

\begin{eqnarray}
\gamma_1^{[1]} & = & x = (\gamma_1^{[3]})^{-1}, \nonumber\\
\gamma_2^{[1]} & = & y = (\gamma_1^{[3]})^{-1} (\gamma_2^{[3]})^{-1}, \nonumber\\
\gamma_1^{[2]} & = & x^{-1}, \nonumber\\
\gamma_2^{[2]} & = & x^{-1} y, \nonumber\\
\gamma_1^{[3]} & = & x^{-1} = \gamma_1^{[2]}, \nonumber\\
\gamma_2^{[3]} & = & x y^{-1} = (\gamma_2^{[2]})^{-1}.
\end{eqnarray}
The Jacobian matrices (note that I am not dealing with inverse Jacobians) for $U_1 \to U_2 \to U_3 \to U_1$ are given by
\begin{eqnarray}
(g^a_i)_{ia} & = & \frac{\partial \gamma_a^{[2]}}{\partial \gamma_i^{[1]}} \nonumber\\
& = & \left( \begin{array}{cc} -x^{-2} & - x^{-2} y \\ 0 & x^{-1} \end{array} \right), \nonumber\\
(g'^{\alpha}_a)_{a \alpha} & = & \frac{\partial \gamma_{\alpha}^{[3]}}{\partial \gamma_a^{[2]}} \nonumber\\
& = & \left( \begin{array}{cc} 1 & 0 \\ 0 & - x^2 y^{-2} \end{array} \right), \nonumber\\
(g''^I_{\alpha})_{\alpha i} & = & \frac{\partial \gamma_I^{[1]}}{\partial \gamma_{\alpha}^{[3]}} \nonumber\\
& = & \left( \begin{array}{cc} -x^2 & - x y \\ 0 & - x^{-1} y^2 \end{array} \right). \nonumber
\end{eqnarray}
Then
\begin{eqnarray}
dg \wedge dg' & = & dx \wedge dy \left( \begin{array}{cc} 0 & 2 x^{-1} y^{-2} \\ 0 & -2 y^{-3} \end{array} \right), \nonumber\\
\psi^0_{123} = tr g'' dg \wedge dg' & = & 2 \frac{dx \wedge dy}{xy}. \nonumber
\end{eqnarray}
Recall that, I defined the last term $\psi^0_{\alpha \beta \gamma}$ of anomaly $2$-form $\psi_{\alpha \beta \gamma}$ (\ref{Psi}) as
\begin{eqnarray}
\psi^0_{\alpha \beta \gamma} = tr g'' dg \wedge d g'. \nonumber
\end{eqnarray}
Similarly, for $U_1 \to U_2 \to U_4 \to U_1$, by the Jacobians $h^a_i, h'^p_a, h''^I_p$, one obtains
\begin{eqnarray}
dh & = & dx \left( \begin{array}{cc} 2 x^{-3} & 2 x^{-3} y \\ 0 & - x^{-2} \end{array} \right) + dy \left( \begin{array}{cc} 0 & x^{-2} \\ 0 & 0 \end{array} \right), \nonumber\\
dh' & = & dx \left( \begin{array}{cc} -2x & y^{-1} \\ 0 & - y^{-2} \end{array} \right) + dy \left( \begin{array}{cc} 0 & -xy^{-2} \\ 0 & 2 x y^{-3} \end{array} \right). \nonumber
\end{eqnarray}
Therefore
\begin{eqnarray}
dh \wedge dh' & = & dx \wedge dy \left( \begin{array}{cc} 0 & x^{-2} y^{-2} \\ 0 & 2 x^{-1} y^{-3} \end{array} \right), \nonumber\\
\psi^0_{124} = tr h'' dh \wedge dh' & = & 2 \frac{dx \wedge dy}{xy}. \nonumber
\end{eqnarray}
For $U_1 \to U_3 \to U_4 \to U_1$, by the Jacobians $k^a_i, k'^p_a, k''^I_p$, one obtains
\begin{eqnarray}
dk & = & dx \left( \begin{array}{cc} 2 x^{-3} & 0 \\ 0 & - y^{-2} \end{array} \right) + dy \left( \begin{array}{cc} 0 & y^{-2} \\ 0 & 2xy^{-3} \end{array} \right), \nonumber\\
dk' & = & dx \left( \begin{array}{cc} -2x & y^{-1} \\ 0 & - x^{-2} \end{array} \right) + dy \left( \begin{array}{cc} 0 & -xy^{-2} \\ 0 & 0 \end{array} \right). \nonumber
\end{eqnarray}
Therefore
\begin{eqnarray}
dk \wedge dk' & = & dx \wedge dy \left( \begin{array}{cc} 0 & -3 x^{-2} y^{-2} \\ 0 & 2 x^{-1} y^{-3} \end{array} \right), \nonumber\\
\psi^0_{134} = tr k'' dk \wedge dk' & = & -2 \frac{dx \wedge dy}{xy}. \nonumber
\end{eqnarray}
For $U_2 \to U_3 \to U_4 \to U_2$, by the Jacobians $l^a_i, l'^p_a, l''^I_p$, one obtains
\begin{eqnarray}
dl & = & dx \left( \begin{array}{cc} 0 & 0 \\ 0 & - 2x y^{-2} \end{array} \right) + dy \left( \begin{array}{cc} 0 & 0 \\ 0 & 2x^2 y^{-3} \end{array} \right), \nonumber\\
dl' & = & dx \left( \begin{array}{cc} -2x & y^{-1} \\ 0 & - x^{-2} \end{array} \right) + dy \left( \begin{array}{cc} 0 & -xy^{-2} \\ 0 & 0 \end{array} \right). \nonumber
\end{eqnarray}
Therefore
\begin{eqnarray}
dl \wedge dl' & = & dx \wedge dy \left( \begin{array}{cc} 0 & 0 \\ 0 & 2 y^{-3} \end{array} \right), \nonumber\\
\psi^0_{234} = tr l'' dl \wedge dl' & = & -2 \frac{dx \wedge dy}{xy}. \nonumber
\end{eqnarray}

\subsection{Antisymmetric $\mu$-term for $1$ point blowup of $\mathbb{CP}^2$}

In the previous section, one obtained the last term $\psi^0_{\alpha \beta \gamma}$ in the anomaly $2$-form (\ref{Psi}) as
\begin{eqnarray}
\psi^0_{\alpha \beta \gamma} & = & c_{\alpha \beta \gamma} \frac{dx \wedge dy}{xy}\mid_{U_{\alpha} \cap U_{\beta} \cap U_{\gamma}}, \label{coefficient}
\end{eqnarray}
where the coefficients $c_{\alpha \beta \gamma}$ are given by
\begin{eqnarray}
c_{123} = 2, \quad
c_{124} = 2, \quad
c_{134} = -2, \quad
c_{234} = -2. \nonumber
\end{eqnarray}
In this section, I discuss the other three $\mu$-terms in (\ref{Psi}).
If $\psi^0_{\alpha \beta \gamma}$ is cancelled by choosing appropriate $\mu$-terms, the anomaly is absent.
This is indeed the case for the $1$ point blowup of $\mathbb{CP}^2$.

This is well described in terms of \v{C}ech cohomology.
\begin{eqnarray}
H^2 (X, \Omega^{cl}) = \frac{Ker (\delta: C^2 \to C^3)}{Im (\delta: C^1 \to C^2)}, \nonumber
\end{eqnarray}
where $X$ is the $1$ point blowup of $\mathbb{CP}^2$ and $\Omega^{cl}$ is the chiral de Rham complex and $C^i$ is $\Omega^{cl}$'s $i$-th \v{C}ech complex and $\delta$ is the corresponding coboundary operator.
$\psi^0_{\alpha \beta \gamma}$ is described as the $2$-cochain $c^2$.
\begin{eqnarray}
(c^2)_{U_i \cap U_j \cap U_k} = c_{ijk}. \nonumber
\end{eqnarray}
The $\mu$-terms are $1$-cochain.
Here I only consider $\mu_{\alpha \beta}$ in the form of
\begin{eqnarray}
\mu_{\alpha \beta} = c_{\alpha \beta} \frac{dx \wedge dy}{xy}. \nonumber
\end{eqnarray}
$\mu_{\alpha \beta}$ must be regular in $U_{\alpha} \cap U_{\beta}$.
By this condition, the variables other than $c_{13}$ or $c_{24}$ are zero.
This can be shown as follows.
We denote the general form of the generator as follows
\begin{eqnarray}
x^L y^M dx \wedge dy, \ (L, M \in \mathbb{Z}), \label{LM}
\end{eqnarray}
where $x$ and $y$ are the affine coordinates of $U_1$.
For the $2$-form (\ref{LM}) to be regular in the patch $U_1$, the integers $L$ and $M$ must satisfy
\begin{eqnarray}
n \ge 0, \quad m \ge 0. \quad (U_1) \nonumber
\end{eqnarray}
In order to determine the regularity condition for $U_2$, I first rewrite the $2$-form (\ref{LM}) with the affine coordinates $(x', y')$ in the patch $U_2$.
The coordinate transformation between $U_1$ and $U_2$ is given in the (Figure \ref{Del1}).
\begin{eqnarray}
x' = x^{-1}, \quad y' = x^{-1} y. \nonumber
\end{eqnarray}
By this coordinate change, the $2$-form (\ref{LM}) is transformed as
\begin{eqnarray}
- x'^{-L-M-3} y'^M dx' \wedge dy'. \nonumber
\end{eqnarray}
The regularity in $U_2$ require that the powers of $x'$ and $y'$ are non-negative, and one obtains the condition
\begin{eqnarray}
M \ge 0, \quad -L - M - 3 \ge 0. \quad (U_2) \nonumber
\end{eqnarray}
In the same way, the condition for $U_3$ and $U_4$ are obtained as follows
\begin{eqnarray}
& & -M - 2 \ge 0, \quad -L- M - 3 \ge 0, \quad (U_3) \nonumber\\
& & L \ge 0, \quad -M -2 \ge 0. \quad (U_4) \nonumber
\end{eqnarray}
These conditions define the allowed region in the lattice ($L$, $M$) for each patch.
The allowed region for $U_i \cap U_j$ can be obtained by the convex hull of two regions for $U_i$ and $U_j$.
We can easily see that the allowed regions for $\displaystyle{U_1 \cap U_2, U_1 \cap U_4, U_2 \cap U_3}$, $\displaystyle{U_3 \cap U_4}$ do not contain $(L, M) = (-1, -1)$, and the corresponding coefficients $c_{\alpha \beta}$ are zero.

The general form of $C^1$ is
\begin{eqnarray}
c^1 & := & c_{13} \frac{dx \wedge dy}{xy}\mid_{U_1 \cap U_3} \oplus c_{24} \frac{dx \wedge dy}{xy}\mid_{U_2 \cap U_4}. \nonumber
\end{eqnarray}
The condition for anomaly cancellation $\psi_{\alpha \beta \gamma} = 0$ is rewritten as
\begin{eqnarray}
c^2 = \delta c^1, \nonumber
\end{eqnarray}
where $\delta$ is the coboundary operator.
This is equivalent to the following relations among coefficients
\begin{eqnarray}
c_{123} & = & c_{23} - c_{13} + c_{12}, \nonumber\\
c_{124} & = & c_{24} - c_{14} + c_{12}, \nonumber\\
c_{134} & = & c_{13} - c_{14} + c_{34}, \nonumber\\
c_{234} & = & c_{23} - c_{24} + c_{34}, \nonumber
\end{eqnarray}
where
\begin{eqnarray}
c_{12} = c_{14} = c_{23} = c_{34} = 0. \nonumber
\end{eqnarray}

Because these equations have the solution
\begin{eqnarray}
c_{13} = -2, \quad c_{24} = 2, \nonumber
\end{eqnarray}
we conclude that
\begin{eqnarray}
c^2 \in Im (\delta: C^1 \to C^2). \nonumber
\end{eqnarray}
Therefore the Pontryagin anomaly vanishes in this case. In the latter chapters, I will compute the total ``gauge-invariant" (namely, no variable-dependent linear combination for the triangulation of the coordinate changing chart).
\begin{eqnarray}
c_{123} + c_{134} & = & c_{124} + c_{234} \nonumber\\
& = & 0. \label{TotalAnomaly1}
\end{eqnarray}

\subsection{Anomaly $2$-forms for $2$ point blowups of $\mathbb{CP}^2$}
For $U_1 \to U_2 \to U_3 \to U_1$, by the Jacobians $g, g', g''$, one obtains
\begin{eqnarray}
dg \wedge dg' & = & dx \wedge dy \left( \begin{array}{cc} 2 x^{-3} & 0 \\ 0 & 0 \end{array} \right), \nonumber\\
\psi^0_{123} := tr g'' dg \wedge dg' & = & - 2 dx \wedge dy x^{-1} y^{-1}, \nonumber\\
c_{123} & = & -2, \nonumber
\end{eqnarray}
where I used the coefficients $c_{\alpha \beta \gamma}$ (\ref{coefficient}) as it was in the last section.
\begin{eqnarray}
\psi^0_{\alpha \beta \gamma} = c_{\alpha \beta \gamma} \frac{dx \wedge dy}{xy}\mid_{U_{\alpha} \cap U_{\beta} \cap U_{\gamma}}. \nonumber
\end{eqnarray}
For $U_1 \to U_2 \to U_4 \to U_1$, by the Jacobians $h, h', h''$, one obtains
\begin{eqnarray}
dh \wedge dh' & = & \left( \begin{array}{cc} 2x^{-1} y^{-2} & 0\\ 0 & 0 \end{array} \right) dx \wedge dy, \nonumber\\
tr h'' dh \wedge dh' & = & 0, \nonumber\\
c_{124} & = & 0. \nonumber
\end{eqnarray}
In the same way, one obtains 
\begin{eqnarray}
& & c_{134} = 1, c_{234} = -1, c_{125} = 0, c_{135} = 0, \nonumber\\
& & c_{145} = -2, c_{235} = -2, c_{245} = -2, c_{345} = -1. \nonumber
\end{eqnarray}
For $\mu_{\alpha \beta} = c_{\alpha \beta} dx \wedge dy / xy$ to be regular in the intersections $U_i \cap U_j$, only the coefficients $c_{13}, c_{14}, c_{24}, c_{25}$, and $c_{35}$ can be non-vanishing. In this case, $\displaystyle{\psi^0_{\alpha \beta \gamma}}$ cannot be cancelled by the $\mu$-terms.
\begin{figure}[htb]
\begin{center}
\includegraphics[width=6cm]{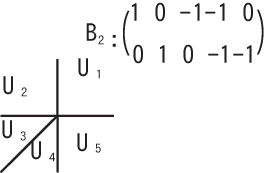}
\end{center}
\caption{Toric diagram for $2$ point blowups of $\mathbb{CP}^2$}
\end{figure}

\begin{figure}[htb]
\begin{center}
\includegraphics[width=9cm]{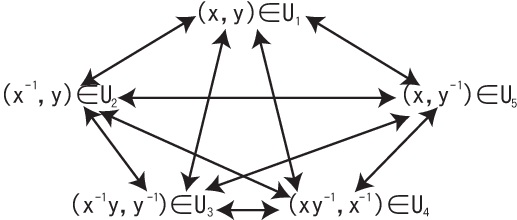}
\end{center}
\caption{Coordinate changes for $2$ point blowups of $\mathbb{CP}^2$}
\label{Del2}
\end{figure}

Namely, there is no solution to the equation
\begin{eqnarray}
c_{\alpha \beta \gamma} = c_{\alpha \beta} + c_{\beta \gamma} + c_{\gamma \alpha}. \label{c_alphabetagamma}
\end{eqnarray}
The absence of the solution can be checked by making the total anomaly, which is the only gauge invariant linear combination of the $2$ cocycles $c_{\alpha \beta \gamma}$.
In this case, the total anomaly is $c_{123}+c_{134}+c_{145}=-2+1-2=-3$.
If one substitutes (\ref{c_alphabetagamma}) into this total anomaly, all the non-vanishing $c_{\alpha \beta}$ cancel, and there is no solution to $c^2 = \delta c^1$.

In general, the total anomaly is given as the sum of $c_{\alpha \beta \gamma}$ for each triangle in a triangulation of the coordinate change diagram (Figure \ref{Del2}). The other cases of ``gauge-invariant" total anomaly are written as follows.
\begin{eqnarray}
c_{123} + c_{134} + c_{145} & = & c_{123} + c_{135} + c_{345} \nonumber\\
& = & c_{124} + c_{234} + c_{145} \nonumber\\
& = & c_{125} + c_{234} + c_{245} \nonumber\\
& = & c_{125} + c_{235} + c_{345} \nonumber\\
& = & -3. \label{TotalAnomaly2}
\end{eqnarray}
\subsection{Anomaly $2$-forms for $3$ point blowups of $\mathbb{CP}^2$}
In this section, I discuss the anomaly for the generic $3$ point blowup of $\mathbb{CP}^2$.
In the same way in the previous sections, I obtain the following coefficients $c_{\alpha \beta \gamma}$ (\ref{coefficient}) of $\psi^0_{\alpha \beta \gamma}$ (\ref{Psi}).
\begin{eqnarray}
& & c_{123} = -1, \quad c_{124} = -2, \quad c_{125} = -2, \quad c_{126} = -1, \nonumber \\
& & c_{134} = -2, \quad c_{135} = -3, \quad c_{136} = -2, \quad c_{145} = -2, \nonumber \\
& & c_{146} = -2, \quad c_{156} = -1, \quad c_{234} = -1, \quad c_{235} = -2, \nonumber \\
& & c_{236} = -2, \quad c_{245} = -2, \quad c_{246} = -3, \quad c_{256} = -2, \nonumber \\
& & c_{345} = -1, \quad c_{346} = -2, \quad c_{356} = -2, \quad c_{456} = -1. \nonumber
\end{eqnarray}
For the intersections $U_i \cap U_j$, only the coefficients $c_{13}, c_{14}, c_{15}, c_{24}, c_{25}, c_{26},$\\
$c_{35}, c_{36},$ and $c_{46}$ are non-vanishing.
\begin{figure}[hbt]
\begin{center}
\includegraphics[width=6cm]{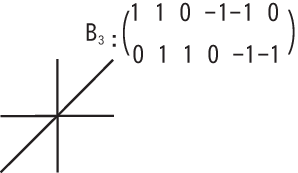}
\end{center}
\caption{Toric diagram for $3$ point blowups of $\mathbb{CP}^2$}
\end{figure}
\begin{figure}[hbt]
\begin{center}
\includegraphics[width=9cm]{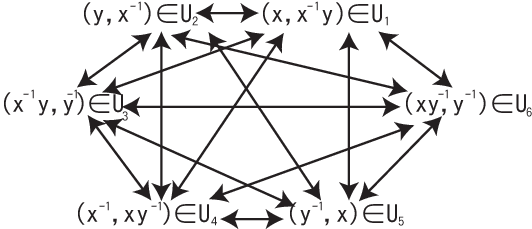}
\end{center}
\caption{Coordinate changes for $3$ point blowups of $\mathbb{CP}^2$}
\label{Del3}
\end{figure}
The total anomaly, the ``gauge-invariant" linear combination of $c_{\alpha \beta \gamma}$, is
\begin{eqnarray}
c_{123}+c_{134}+c_{145}+c_{156} = -6. \nonumber
\end{eqnarray}
Likewise, we can compute all of the possible triangulations of the coordinate changing charts as follows. (The order is in cyclic and dictionary-like numbers.)
\begin{eqnarray}
c_{123} + c_{134} + c_{145} + c_{156} & = & c_{123} + c_{134} + c_{146} + c_{456} \nonumber\\
& = & c_{123} + c_{135} + c_{345} + c_{156} \nonumber\\
& = & c_{123} + c_{136} + c_{346} + c_{456} \nonumber\\
& = & c_{123} + c_{136} + c_{356} + c_{345} \nonumber\\
& = & c_{124} + c_{234} + c_{145} + c_{156} \nonumber\\
& = & c_{124} + c_{234} + c_{146} + c_{456} \nonumber\\
& = & c_{125} + c_{234} + c_{245} + c_{156} \nonumber\\
& = & c_{125} + c_{235} + c_{345} + c_{156} \nonumber\\
& = & c_{126} + c_{234} + c_{245} + c_{256} \nonumber\\
& = & c_{126} + c_{234} + c_{246} + c_{456} \nonumber\\
& = & c_{126} + c_{235} + c_{345} + c_{256} \nonumber\\
& = & c_{126} + c_{236} + c_{346} + c_{456} \nonumber\\
& = & c_{126} + c_{236} + c_{356} + c_{345} \nonumber\\
& = & -6. \label{TotalAnomaly3}
\end{eqnarray}
From the results of section $3.2, 4.2, 4.3, 4.4$, I find that the total anomaly with the proportional constant $-1$ (the minus sign from the definition of $\psi^0_{\alpha \beta \gamma}$ inside $\psi_{\alpha \beta \gamma}$ (\ref{Psi})) is
\begin{eqnarray}
3 (n-1), \nonumber
\end{eqnarray}
for $n=0,1,2,3$ generic point blowup of $\mathbb{CP}^2$ (\ref{TotalAnomaly0})(\ref{TotalAnomaly1})(\ref{TotalAnomaly2})(\ref{TotalAnomaly3}) (toric del Pezzo surfaces).
The second Chern character by the Riemann-Roch theorem (\ref{RiemannRoch}) is in agreement with this discussion.

\section{Conclusion and future direction}

In this paper, I explicitly examined by $2$ ways that the Ans\"{a}tze of Witten's
heterotic $\mathcal{N}=(0,2)$ model and Nekrasov's generalized complex geometry are consistent. One way is by step by step careful OPE calculation and the other is the computation of the anomaly $2$-form -- the $2$-cocycle of the chiral de Rham complex -- in terms of coordinate transformation Jacobian matrices. I computed the anomaly $2$-forms in the case of toric del Pezzo surfaces of all degrees and conclude that this coincides with the results (\ref{RiemannRoch}) of Riemann-Roch theorem.

Notice that Beilinson-Drinfeld chiral algebra has a background of geometric Langlands programs studied on the geometric quantization of symplectic manifolds (stable Higgs bundles\cite{Simpson} satisfying the self-duality equation of Hitchin over a Riemann surface), which is similar to physicists' theory of topological fields or topological strings as quantization of $2$-dimensional Yang-Mills-Higgs theory.
This is a bridge between the twist methods of $4$ dimensional ($\mathcal{N} = 4$ hyperk\"ahler) or $6$ dimensional ($\mathcal{N} = 2$ K\"ahler) Super-Yang-Mills theory of Gukov\cite{GW}, Kapustin-Witten\cite{KW} et al. and the local / global (quantum) geometric Langlands program of Edward Frenkel, Gaitsgory\cite{FG} et al.

I will also try to understand the gerbes of chiral differential operators in the language of stability conditions for $2$-category theory demanded from the chiral de Rham complex on stacks and the chiral homology theory of algebro-geometrized conformal blocks for the existence of global section of twisted $\mathcal{D}$-modules.  It is elusive whether or not one can physically and geometrically realize the idea of recent topological $\epsilon$-factor theory (analogue of functional equation of $L$-function) of Beilinson\cite{B} after the differential graded algebra (DGA) theory for rational homotopy or motivic homology (or rather, motivic homotopy).

In addition, I hope to understand the Kashiwara(-Schapira-\\
Schneider) type index theorem for world-sheet with boundary;
rather than a generalized Atiyah-Singer index theorem for family of Dirac operators on free loop space of non-compact or with boundary
of either special Lagrangian submanifolds or coisotropic submanifolds of K\"ahler stack.
In this sense, the compactification of stable Higgs bundle of real $4$-dimensional super Yang-Mills theory on Riemann surface $\Sigma$
(having another Riemann surface $C$ as a fiber) is not yet settled.
\newpage
This is because, in the mathematics literature, the $G$-torsor $Bun_G (C)$ of reductive algebraic group $G$ will be the target space of sigma model having maps from the world-sheet $\Sigma$;
whereas, in the physics literature, it seems that, $Bun_G (C)$ denotes the moduli space of triple $(A, F_A, \Phi)$, where $A$ is a connection of the principal $G$-bundle over $C$, $F_A$ is the curvature of this bundle, and $\Phi$ is a $(1,0)$-form with a $(0,1)$-form of Higgs field as the Hermitian-Yang-Mills theory in addition to just a gauge theory of principal $G$-bundles.
\begin{eqnarray}
F \pm \Phi \wedge \bar{\Phi} & = & 0, \\
\bar{\nabla_{A}} \phi & = & \nabla_A \bar{\phi} \nonumber\\
& = & 0, \nonumber\\
\Phi & = & \phi dz + \bar{\phi} d \bar{z} = \bar{\Phi}.
\end{eqnarray}

Moreover, the model of physicists seems to be a topological sigma model
of several kinds (say, A-, B-, I-) from world-sheet $\Sigma$ to
hyperk\"ahler stack $T^{\ast} Bun_G (C)$, whose dimension is twice the dimension
of the definition of some of the mathematics literature.
It is in discussion whether the Lagrangian submanifolds, whose real
dimension is half the real dimension of the ``target space" as either $Bun_G (C)$
or $T^{\ast} Bun_G (C)$, are odd or even dimensional real submanifolds;
namely, whether these admit complex structures.
This is the difficult point to the definition of object of the Fukaya category of the target orbifold.

During the preparation of this paper, Kapustin-Witten\cite{KW} submitted a paper to the preprint, which is related to this paper.

\appendix

\section{Wess-Zumino-Witten term}
The Wess-Zumino-Witten term is used as a term corresponding to the antisymmetric $\mu$ term, which is used in the Nekrasov's Ansatz (\ref{NekrasovAnsatz}).
I will make a historical \footnote{See \cite{APW} and related papers for physical discussions. I did not cite the very original papers that were published more than $30$ years ago.}note on this term.
First, I will think of the $SU(3)_L \times SU(3)_R$ spontaneously broken to the diagonal $SU(3)$, which involves the Nambu-Goldstone boson $\pi$.
\begin{eqnarray}
\mathcal{L} & = & \frac{1}{16 \pi} F_{\pi}^2 \int d^4 x Tr \partial_\mu U \partial_\mu U^{-1}, \nonumber
\end{eqnarray}
where $F_{\pi}$ is an undetermined real constant. Then, the Euler-Lagrange equation is
\begin{eqnarray}
\partial_{\mu} (\frac{1}{8} F_{\pi}^2 U^{-1} \partial_{\mu} U) & = & 0. \nonumber
\end{eqnarray}
I will add a term, which violate parity $P_0$ that changes $x \to -x$, $t \to t$, $U \to U$.
\begin{align}
& \partial_{\mu} (\frac{1}{8} F_{\pi}^2 U^{-1} \partial_{\mu} U) + \lambda \epsilon^{\mu \nu \alpha \beta} U^{-1} (\partial_{\mu} U) U^{-1} (\partial_{\nu} U) U^{-1} (\partial_{\alpha} U) U^{-1} (\partial_{\beta} U)\nonumber\\
= & 0, \nonumber
\end{align}
where the second term is with undetermined real constant $\lambda$ and antisymmetric tensor $\epsilon^{\mu \nu \alpha \beta}$ (Edington's symbol). I would like to derive this equation from a Lagrangian, which is difficult in the first sight.

\subsection{Analogy with particle of magnetic monopoles}
The equation of motion in the constrained system $\Sigma x_i^2 = 1$ can have both $x \to -x$ and $t \to -t$ symmetry, if we write it as follows
\begin{eqnarray}
m \frac{\partial^2 x_i}{\partial t^2} + m x_i (\sum_k (\frac{\partial x_k}{dt})^2) & = & \alpha \epsilon_{ijk} x_j \frac{\partial x_k}{\partial t}, \nonumber
\end{eqnarray}
where $\alpha$ is an undetermined real constant and $\epsilon$ is the antisymmetric tensor.
As the right hand side can be seen as the Lorentz force for an electric charge interacting with a magnetic monopole located at the center of the sphere. In this case the vector potential is, by definition,
\begin{eqnarray}
\nabla \times \vec{A} & = & \frac{\vec{x}}{|x|^3}. \nonumber
\end{eqnarray}
Then by the Stokes theorem, we can rewrite the contour integral of vector potential as the integration of field strength (flux) trough a topological disk $D$, whose choice can be arbitrary and we can especially take as $D'$ (orientation reversed).
\begin{eqnarray}
1 & = & \exp (\sqrt{-1} \alpha \int_{D+D'} F_{ij} d \Sigma^{ij}), \nonumber
\end{eqnarray}
where $d \Sigma^{ij}$ is a cubic unit area element of sphere. Therefore $\alpha$ is integer or half-integer, which is the Dirac quantization.

\subsection{Recovery of the equation of motion of\\Wess-Zumino-Witten model}
As is noted above, we need some term proportional to the field strength, so we let $Q$ be five-dimensional disc which has the $SU(3)$ as the boundary because of $\pi_4(SU(3)) = 0$
\begin{eqnarray}
\Gamma & = & \int_Q \omega_{ijklm} d \Sigma^{ijklm}. \nonumber
\end{eqnarray}
$Q$ can be deformed to orientation revered $Q'$, then $Q+Q'=S$ is a five-dimensional sphere $S^5$.
\begin{eqnarray}
\int_S \omega_{ijklm} d \Sigma^{ijklm} & = & 2 \pi \cdot integer. \nonumber
\end{eqnarray}
Then $S$ is in $SU(3)$ and $\pi_5(SU(3)) = \mathbb{Z}$ therefore, we can take a unit sphere $S_0$ and write the action $I$ as
\begin{eqnarray}
I & = & \frac{1}{16 \pi} F_{\pi}^2 \int d^4x Tr \partial_{\mu} U \partial_{\mu} U^{-1} + n \Gamma. \nonumber
\end{eqnarray}

\section{Toric diagrams and birational geometry}
Toric varieties are algebraic manifolds which have the action of algebraic torus
$(\mathbb{C}^{\times})^n = (\mathbb{C} \setminus \{0\})^n$, where $n$ is the
complex dimension of the varieties.
Let us consider the case of $n=2$.
For each toric diagram (fan), we can define
the following data. Vertices $e_i \in \mathbb{Z}^2$, and vectors $v_i$ from the origin to the vertices.
The dual basis $w_i^1, w_i^2$ for affine patch $U_i$ spanned by the $2$ vectors $v_i, v_{i+1}$
is as follows:
\begin{eqnarray}
(v_i, w_i^1) = 1, \quad (v_i, w_i^2) = 0, \quad (v_{i+1}, w_i^1) = 0, \quad (v_{i+1}, w_i^2) = 1. \nonumber
\end{eqnarray}
The affine patch $U_i$ is described by
\begin{eqnarray}
Spec \mathbb{C} [x^{(w_i^1, E_1)}y^{(w_i^1, E_2)}, x^{(w_i^2, E_1)} y^{(w_i^2, E_2)}], \nonumber
\end{eqnarray}
where $E_1 = (1,0)$, and $E_2 = (0,1)$ are the standard bases of $\mathbb{Z}^2$.
Namely the canonical coordinates $v^{[i]}, w^{[i]}$ in the affine patch $U_i$ are given by
\begin{eqnarray}
v^{[i]} & = & x^{(w_i^1, E_1)}y^{(w_i^1, E_2)}, \nonumber\\
w^{[i]} & = & x^{(w_i^2, E_1)} y^{(w_i^2, E_2)}. \nonumber
\end{eqnarray}
The \footnote{Conventional definition and not canonical in the categorical sense.}blowups are operations in the divisor linear system, where we assign a
new vertex $e_{n+1} \in \mathbb{Z}^2$ in the generic point and the corresponding
vector from the origin. This operation is birational since we have the toric action on the toric variety, which includes the
toric variety $(\mathbb{C}^{\times})^2$ as the dense submanifolds and therefore
the toric diagram in $\mathbb{Z}^2$ has the $GL(2)$ action.

\textit{E-mail address}\\
\hspace*{8mm}
Makoto Sakurai: makotosakuraijp@08.alumni.u-tokyo.ac.jp
\vskip 0.5cm
Graduate School of Mathematical Sciences,\\
The University of Tokyo, 3-8-1 Komaba,\\
Meguro, Tokyo 153-8914, Japan (Alumnus)
\vskip 0.5cm
Tokyo University of Technology, (Year 2011-2012)\\
1404-1 Katakuramachi, Hachioji City, Tokyo 192-0982, Japan.

\end{document}